\documentclass[notitlepage,12pt]{article}
\usepackage{psfig}

\begin{document}

\title{Why does a metal--superconductor junction have a resistance?}
\author{C. W. J. Beenakker
\\
Instituut-Lorentz, Universiteit Leiden\\
P.O. Box 9506, 2300 RA Leiden, The Netherlands}
\date{September 1999}

\maketitle

\begin{abstract}
This is a tutorial article based on a lecture delivered
in June 1999 at the NATO Advanced Study Institute in Ankara.
The phenomenon of Andreev reflection is introduced as the electronic analogue
of optical phase-conjugation. In the optical problem, a disordered medium
backed by a phase-conjugating mirror can become completely transparent. Yet, a
disordered metal connected to a superconductor has the same resistance as in
the normal state. The resolution of this paradox teaches us a fundamental
difference between phase conjugation of light and electrons.\medskip\\
To be published in {\em Quantum Mesoscopic Phenomena and
Mesoscopic Devices in Microelectronics}, edited by I. O. Kulik
and R. Ellialtioglu (Kluwer, Dordrecht).
\end{abstract}

\section{Introduction}

In the late sixties, Kulik used the mechanism of Andreev reflection
\cite{And64} to explain how a metal can carry a dissipationless current between
two superconductors over arbitrarily long length scales, provided the
temperature is low enough \cite{Kul69}. One can say that the normal metal has
become superconducting because of the proximity to a superconductor. This
proximity effect exists even if the electrons in the normal metal have no
interaction. At zero temperature the maximum supercurrent that the metal can
carry decays only algebraically with the separation between the superconductors
--- rather than exponentially, as it does at higher temperatures.

The recent revival of interest in the proximity effect has produced a deeper
understanding into how the proximity-induced superconductivity of
non-interacting electrons differs from true superconductivity of electrons
having a pairing interaction. Clearly, the proximity effect does not require
two superconductors. One should be enough. Consider a junction between a normal
metal and a superconductor (an NS junction). Let the temperature be zero. What
is the resistance of this junction? One might guess that it should be smaller
than in the normal state, perhaps even zero. Isn't that what the proximity
effect is all about?

The answer to this question has been in the literature since 1979 \cite{Art79},
but it has been appreciated only in the last few years. A recent review
\cite{Cou99} gives a comprehensive discussion within the framework of the
semiclassical theory of superconductivity. A different approach, using
random-matrix theory, was reviewed by the author \cite{Bee97}. In this lecture
we take a more pedestrian route, using the analogy between Andreev reflection
and optical phase-conjugation \cite{Len90,Hou91} to answer the question: Why
does an NS junction have a resistance?

\section{Andreev reflection and optical phase-conjugation}

\begin{figure}[tb]
\centerline{\psfig{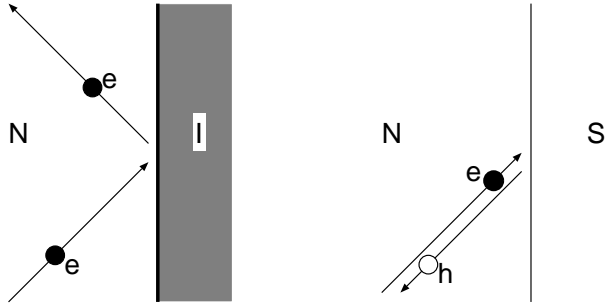}\medskip\\}
\caption[]{
Normal reflection by an insulator (I) versus Andreev reflection by a
superconductor (S) of an electron excitation in a normal metal (N) near the
Fermi level. Normal reflection (left) conserves charge but does not conserve
momentum. Andreev reflection (right) conserves momentum but does not conserve
charge: The electron (e) is reflected as a hole (h) with the same momentum and
opposite velocity. The missing charge of $2e$ is absorbed as a Cooper pair by
the superconducting condensate.
\label{reflection}}
\end{figure}

It was first noted by Andreev in 1963 \cite{And64} that an electron is
reflected from a superconductor in an unusual way. The differences between
normal reflection and Andreev reflection are illustrated in Fig.\
\ref{reflection}. Let us discuss them separately.
\begin{itemize}
\item
{\em Charge is conserved in normal reflection but not in Andreev reflection.}
The reflected particle (the hole) has the opposite charge as the incident
particle (the electron). This is not a violation of a fundamental conservation
law. The missing charge of $2e$ is absorbed into the superconducting ground
state as a Cooper pair. It is missing only with respect to the excitations.
\item
{\em Momentum is conserved in Andreev reflection but not in normal reflection.}
The conservation of momentum is an approximation, valid if the superconducting
excitation gap $\Delta$ is much smaller than the Fermi energy $E_{\rm F}$ of
the normal metal. The explanation for the momentum conservation is that the
superconductor can not exert a significant force on the incident electron,
because $\Delta$ is too small compared to the kinetic energy $E_{\rm F}$ of the
electron \cite{Abr88}. Still, the superconductor has to reflect the electron
somehow, because there are no excited states within a range $\Delta$ from the
Fermi level. It is the unmovable rock meeting the irresistible object. Faced
with the challenge of having to reflect a particle without changing its
momentum, the superconductor finds a way out by transforming the electron into
a particle whose velocity is opposite to its momentum: a hole.
\item
{\em Energy is conserved in both normal and Andreev reflection.} The electron
is at an energy $\varepsilon$ above the Fermi level and the hole is at an
energy $\varepsilon$ below it. Both particles have the same excitation energy
$\varepsilon$. Andreev reflection is therefore an {\em elastic\/} scattering
process.
\item
{\em Spin is conserved in both normal and Andreev reflection.} To conserve
spin, the hole should have the opposite spin as the electron. This spin-flip
can be ignored if the scattering properties of the normal metal are
spin-independent.
\end{itemize}

\begin{figure}[tb]
\centerline{\psfig{file=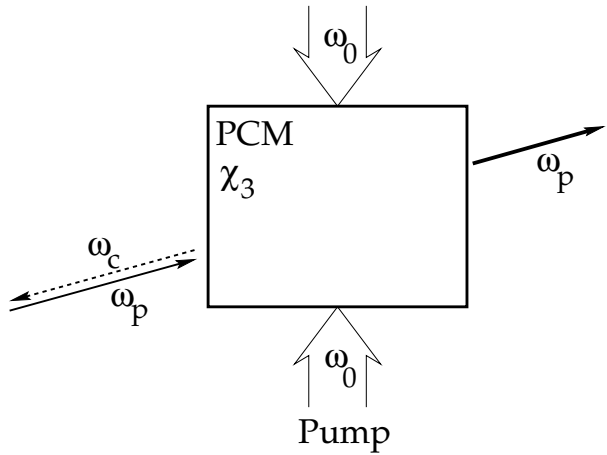,width= 8cm,angle=-90}\medskip\\}
\caption[]{
Schematic drawing of optical phase-conjugation by means of four-wave mixing.
The phase-conjugating mirror (PCM) consists of a cell filled by a medium with a
third-order non-linear susceptibility $\chi_{3}$. (Examples are ${\rm
BaTiO}_{3}$ and ${\rm CS}_{2}$.)
The medium is pumped by two counter-propagating beams at frequency
$\omega_{0}$. A probe beam incident at frequency $\omega_{\rm
p}=\omega_{0}+\delta\omega$ is then retro-reflected as a conjugate beam at
frequency $\omega_{\rm c}=\omega_{0}-\delta\omega$. From Ref.\
\protect\cite{Paa97}.
\label{pcmsys}}
\end{figure}

\noindent
The NS junction has an optical analogue known as a phase-conjugating mirror
\cite{Pep86}. Phase conjugation is the effect that an incoming wave
$\propto\cos(kx-\omega t)$ is reflected as a wave $\propto\cos(-kx-\omega t)$,
with opposite sign of the phase $kx$. Since $\cos(-kx-\omega t)=\cos(kx+\omega
t)$, this is equivalent to reversing the sign of the time $t$, so that phase
conjugation is sometimes called a time-reversal operation. The reflected wave
has a wavevector precisely opposite to that of the incoming wave, and therefore
propagates back along the incoming path. This is called retro-reflection. Phase
conjugation of light was discovered in 1970 by Woerdman and by Stepanov,
Ivakin, and Rubanov \cite{Woe70,Ste71}.

A phase-conjugating mirror for light (see Fig.\ \ref{pcmsys}) consists of a
cell containing a liquid or crystal with a large nonlinear susceptibility. The
cell is pumped by two counter-propagating beams at frequency $\omega_{0}$. A
third beam is incident with a much smaller amplitude and a slightly different
frequency $\omega_{0}+\delta\omega$. The non-linear susceptibility leads to an
amplification of the incident beam, which is transmitted through the cell, and
to the generation of a fourth beam, which is reflected. This non-linear optical
process is called ``four-wave mixing''. Two photons of the pump beams are
converted into one photon for the transmitted beam and one for the reflected
beam. Energy conservation dictates that the reflected beam has frequency
$\omega_{0}-\delta\omega$. Momentum conservation dictates that its wavevector
is opposite to that of the incident beam. Comparing retro-reflection of light
with Andreev reflection of electrons, we see that the Fermi energy $E_{\rm F}$
plays the role of the pump frequency $\omega_{0}$, while the excitation energy
$\varepsilon$ corresponds to the frequency shift $\delta\omega$.

\begin{figure}[tb]
\centerline{\psfig{file=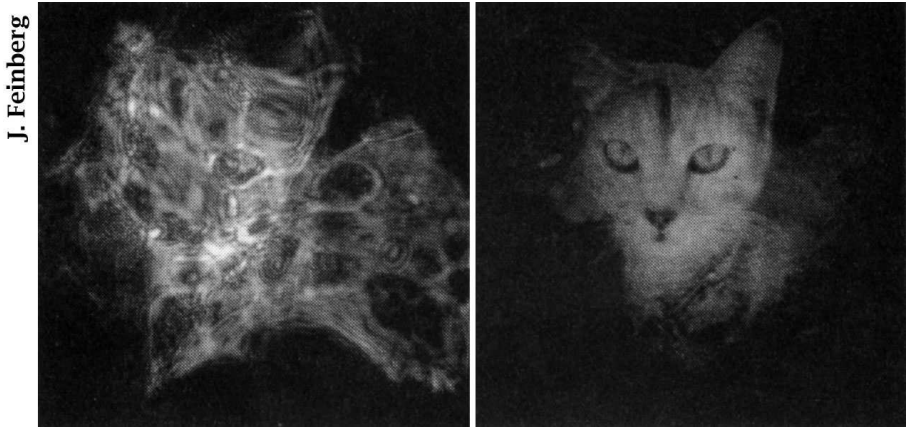,width= 12cm}}
\caption[]{
Example of wavefront reconstruction by optical phase-conjugation. In both
photographs the image of a cat was distorted by transmitting it through a piece
of frosted glass, and reflecting it back through the same piece of glass. This
gives an unrecognizable image when reflected by an ordinary mirror (left panel)
and the original image when reflected by a phase-conjugating mirror (right
panel). From Ref.\ \protect\cite{Fei82}.
\label{cat}}
\end{figure}

A phase-conjugating mirror can be used for wavefront reconstruction. Imagine an
incoming plane wave, that is distorted by some inhomogeneity. When this
distorted wave falls on the mirror, it is phase conjugated and retro-reflected.
Due to the time-reversal effect, the inhomogeneity that had distorted the wave
now changes it back to the original plane wave. An example is shown in Fig.\
\ref{cat}. Complete wavefront reconstruction is possible only if the distorted
wavefront remains approximately planar, since perfect time reversal upon
reflection holds only in a narrow range of angles of incidence for realistic
systems. This is an important, but not essential complication, that we will
ignore in what follows.

\section{The resistance paradox}

\begin{figure}[tb]
\centerline{\psfig{file=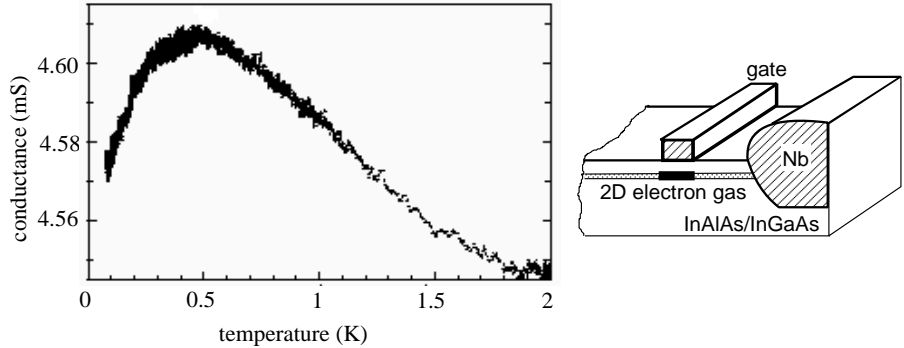,width= 12cm}}
\caption[]{
Temperature dependence of the conductance of an NS junction, showing the
re-entrance effect. The superconductor is Nb, the normal metal is a
two-dimensional electron gas. A gate creates a strongly disordered region in
the 2D gas that dominates the conductance of the junction. Upon lowering the
temperature the conductance first rises and then drops again. Under ideal
circumstances the low- and high-temperature limits would be the same. From
Ref.\ \protect\cite{Toy99}.
\label{reentrant}}
\end{figure}

We have learned that a disordered medium (such as the frosted glass in Fig.\
\ref{cat}) becomes transparent when it is backed by a phase-conjugating mirror.
By analogy, one would expect that a disordered metal backed by a superconductor
would become ``transparent'' too, meaning that its resistance should vanish (up
to a small contact resistance that is present even without any disorder). This
does not happen. Upon decreasing the temperature below the superconducting
transition temperature, the resistance drops slightly but then rises again back
to its high-temperature value. (A recent experiment is shown in Fig.\
\ref{reentrant}, where the conductance is plotted instead of the resistance.)
This socalled ``re-entrance effect'' has been reviewed recently by Courtois
{\em et al.\/} \cite{Cou99}, and we refer to that review for an extensive list
of references. The theoretical prediction \cite{Art79,Bee92,Naz96} is that {\em
at zero temperature the resistance of the normal-metal--superconductor junction
is the same as in the normal state.} How can we reconcile this with the notion
of Andreev reflection as a ``time-reversing'' process, analogous to optical
phase-conjugation? To resolve this paradox, let us study the analogy more
carefully, to see where it breaks down.

\begin{figure}[tb]
\centerline{\psfig{file=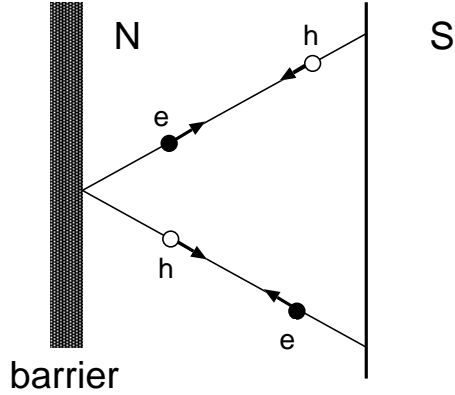,width= 6cm}}
\caption[]{
Periodic orbit consisting of two normal reflections and two retro-reflections.
The net phase increment is zero in the optical case and $-\pi$ in the
electronic case. Hence the periodic orbits interfere constructively for light
and destructively for electrons. This explains why the barrier becomes
transparent for light but not for electrons.
\label{orbit}}
\end{figure}

For a simple discussion it is convenient to replace the disordered medium by a
tunnel barrier (or semi-transparent mirror) and consider the phase shift
accumulated by an electron (or light wave) that bounces back and forth between
the barrier and the superconductor (or phase-conjugating mirror). A periodic
orbit (see Fig.\ \ref{orbit}) consists of two round-trips, one as an electron
(or light at frequency $\omega_{0}+\delta\omega$), the other as a hole (or
light at frequency $\omega_{0}-\delta\omega$). The miracle of phase conjugation
is that phase shifts accumulated in the first round trip are cancelled in the
second round trip. If this were the whole story, one would conclude that the
net phase increment is zero, so all periodic orbits would interfere
constructively and the tunnel barrier would become transparent because of
resonant tunneling.

But it is not the whole story. There is an extra phase shift of $-\pi/2$
acquired upon Andreev reflection that destroys the resonance. Since the
periodic orbit consists of two Andreev reflections, one from electron to hole
and one from hole to electron, and both reflections have the same phase shift
$-\pi/2$, the net phase increment of the periodic orbit is $-\pi$ and not zero.
So subsequent periodic orbits interfere destructively, rather than
constructively, and tunneling becomes suppressed rather than enhanced. In
contrast, a phase-conjugating mirror adds a phase shift that alternates between
$+\pi/2$ and $-\pi/2$ from one reflection to the next, so the net phase
increment of a periodic orbit remains zero.

For a more quantitative description of the conductance we need to compute the
probability $R_{\rm he}$ that an incident electron is reflected as a hole. The
matrix of probability amplitudes $r_{\rm he}$ can be constructed as a geometric
series of multiple reflections:
\begin{eqnarray}
r_{\rm he}&=&t^{\dagger}\frac{1}{\rm i}t+t^{\dagger} \frac{1}{\rm
i}r\frac{1}{\rm i}r^{\dagger}\frac{1}{\rm i}t
+t^{\dagger}\frac{1}{\rm i}\left[r\frac{1}{\rm i}r^{\dagger} \frac{1}{\rm
i}\right]^{2}t+\cdots \nonumber\\
&=&t^{\dagger}\frac{1}{\rm i}\left[1-r\frac{1}{\rm i}r^{\dagger}\frac{1}{\rm
i}\right]^{-1}t. \label{rhe}
\end{eqnarray}
Each factor $1/{\rm i}=\exp(-{\rm i}\pi/2)$ corresponds to an Andreev
reflection. The matrices $t,t^{\dagger}$ and $r,r^{\dagger}$ are the $N\times
N$ transmission and reflection matrices of the tunnel barrier, or more
generally, of the disordered region in the normal metal. (The number $N$ is
related to the cross-sectional area $A$ of the junction and the Fermi
wavelength $\lambda_{\rm F}$ by $N\simeq A/\lambda_{\rm F}^{2}$.) The matrices
$t,r$ pertain to the electron and the matrices $t^{\dagger},r^{\dagger}$ to the
hole. The resulting reflection probability $R_{\rm he}=N^{-1}\,{\rm Tr}\,r_{\rm
he}^{\vphantom{\dagger}}r_{\rm he}^{\dagger}$ is given by \cite{Bee92}
\begin{equation}
R_{\rm he}=\frac{1}{N}\,{\rm
Tr}\,\left(\frac{tt^{\dagger}}{1+rr^{\dagger}}\right)^{2} =\frac{1}{N}\,{\rm
Tr}\,\left(\frac{tt^{\dagger}}{2-tt^{\dagger}}\right)^{2}. \label{Rhe}
\end{equation}
We have used the relationship $tt^{\dagger}+rr^{\dagger}=1$, dictated by
current conservation. The conductance $G_{\rm NS}$ of the NS junction is
related to $R_{\rm he}$ by \cite{Blo82,Tak92}
\begin{equation}
G_{\rm NS}=\frac{4e^{2}}{h}NR_{\rm he}.\label{GNS}
\end{equation}

In the optical analogue one has the probability $R_{\pm}$ for an incident light
wave with frequency $\omega_{0}+\delta\omega$ to be reflected into a wave with
frequency $\omega_{0}-\delta\omega$. The matrix of probability amplitudes is
given by the geometric series
\begin{eqnarray}
r_{\pm}&=&t^{\dagger}\frac{1}{\rm i}t+t^{\dagger} \frac{1}{\rm i}r{\rm
i}r^{\dagger}\frac{1}{\rm i}t
\mbox{}+t^{\dagger}\frac{1}{\rm i}\left[r{\rm i}r^{\dagger}\frac{1}{\rm
i}\right]^{2}t+\cdots \nonumber\\
&=&t^{\dagger}\frac{1}{\rm i}\left[1-r{\rm i}r^{\dagger}\frac{1}{\rm
i}\right]^{-1}t. \label{rpm}
\end{eqnarray}
The only difference with Eq.\ (\ref{rhe}) is the alternation of factors $1/{\rm
i}$ and ${\rm i}$, corresponding to the different phase shifts $\exp(\pm {\rm
i}\pi/2)$ acquired at the phase-conjugating mirror. The reflection probability
$R_{\pm}=N^{-1}\,{\rm Tr}\,r_{\pm}^{\vphantom{\dagger}}r_{\pm}^{\dagger}$ now
becomes independent of the disorder \cite{Paa97b},
\begin{equation}
R_{\pm}=\frac{1}{N}\,{\rm
Tr}\,\left(\frac{tt^{\dagger}}{1-rr^{\dagger}}\right)^{2} =1. \label{Rpm}
\end{equation}
{\em The disordered medium has become completely transparent.}

It is remarkable that a small difference in phase shifts has such far reaching
consequences. Note that one needs to consider multiple reflections in order to
see the difference: The first term in the series is the same in Eqs.\
(\ref{rhe}) and (\ref{rpm}). That is probably why this essential difference
between Andreev reflection and optical phase-conjugation was not noticed prior
to Ref.\ \cite{Paa97b}.

\section{How big is the resistance?}

Now that we understand why a disordered piece of metal connected to a
superconductor does not become transparent, we would like to go one step
further and ask whether the resistance (or conductance) is bigger or smaller
than without the superconductor. To that end we compare, following Ref.\
\cite{Bee92}, the expression for the conductance of the NS junction [obtained
from Eqs.\ (\ref{Rhe}) and (\ref{GNS})],
\begin{equation}
G_{\rm NS}=\frac{4e^{2}}{h}\sum_{n=1}^{N}\frac{T_{n}^{2}}{(2-T_{n})^{2}},
\label{keyzero}
\end{equation}
with the Landauer formula for the normal-state conductance,
\begin{equation}
G_{\rm N}=\frac{2e^{2}}{h}\sum_{n=1}^{N}T_{n}.\label{Landauer}
\end{equation}
The numbers $T_{1},T_{2},\ldots T_{N}$ are the eigenvalues of the matrix
product $tt^{\dagger}$. These transmission eigenvalues are real numbers between
0 and 1 that depend only on the properties of the metal (regardless of the
superconductor). Both formulas (\ref{keyzero}) and (\ref{Landauer}) hold at
zero temperature, so we will be comparing the zero-temperature limits of
$G_{\rm NS}$ and $G_{\rm N}$.

Since $x^{2}/(2-x)^{2}\leq x$ for $x\in[0,1]$, we can immediately conclude that
$G_{\rm NS}\leq 2G_{\rm N}$. If there is no disorder, then all $T_{n}$'s are
equal to unity, hence $G_{\rm NS}$ reaches its maximum value of $2G_{\rm N}$.
For a tunnel barrier all $T_{n}$'s are $\ll 1$, hence $G_{\rm NS}$ drops far
below $G_{\rm N}$. A disordered metal will lie somewhere in between these two
extremes, but where?

We have already alluded to the answer in the previous section, that $G_{\rm
NS}=G_{\rm N}$ for a disordered metal in the zero-temperature limit. To derive
this remarkable equality, we parameterize the transmission eigenvalue $T_{n}$
in terms of the localization length $\zeta_{n}$,
\begin{equation}
T_{n}=\frac{1}{\cosh^{2}(L/\zeta_{n})},\label{xip}
\end{equation}
where $L$ is the length of the disordered region. Substitution into Eqs.\
(\ref{keyzero}) and (\ref{Landauer}) gives the average conductances
\begin{eqnarray}
\langle G_{\rm
NS}\rangle_{L}&=&\frac{4e^{2}}{h}N\int_{0}^{\infty}\!d\zeta\,P_{L}(\zeta)
\cosh^{-2}(2L/\zeta),\label{GNSaverage}\\
\langle G_{\rm
N}\rangle_{L}&=&\frac{2e^{2}}{h}N\int_{0}^{\infty}
\!d\zeta\,P_{L}(\zeta)\cosh^{-2}(L/\zeta). \label{GNaverage}
\end{eqnarray}
(For Eq.\ (\ref{GNSaverage}) we have used that $2\cosh^{2}x-1=\cosh 2x$.) The
probability distribution $P_{L}(\zeta)$ of $\zeta$ is independent of $L$ in a
range of lengths between $l$ and $Nl$ \cite{Bee97}. It then follows immediately
that
\begin{equation}
\langle G_{\rm NS}\rangle_{L}=2\langle G_{\rm N}\rangle_{2L}.\label{GNSGN}
\end{equation}
Since $G_{\rm N}\propto 1/L$, according to Ohm's law, we arrive at the equality
of $G_{\rm NS}$ and $G_{\rm N}$.

The restriction to the range $l\ll L\ll Nl$ is the restriction to the regime of
diffusive transport: For smaller $L$ we enter the ballistic regime and $G_{\rm
NS}$ rises to $2G_{\rm N}$; For larger $L$ we enter the localized regime, where
tunneling takes over from diffusion and $G_{\rm NS}$ becomes $\ll G_{\rm N}$.

\section{Conclusion}

We have learned a fundamental difference between Andreev reflection of
electrons and phase-conjugation of light. While it is appealing to think of the
Andreev reflected hole as the time reverse of the incident electron, this
picture breaks down upon closer inspection. The phase shift of $-\pi/2$
acquired upon Andreev reflection spoils the time-reversing properties and
explains why a disordered metal does not become transparent when connected to a
superconductor.
\bigskip\\
The research on which this lecture is based was done in collaboration with
J.~C.~J. Paasschens. It was supported by the ``Stichting voor Fundamenteel
Onderzoek der Materie'' (FOM) and by the ``Nederlandse organisatie voor
Wetenschappelijk Onderzoek'' (NWO).

\end{document}